\documentclass[twocolumn]{aastex63}
\usepackage{amsmath}
\usepackage{mathtools}
\usepackage{amssymb}
\usepackage{lineno}

\shorttitle{X-ray Plateau of GRB 230307A}
\shortauthors{Zhong et al.}

\begin{document}
\title{The Very Early Soft X-ray Plateau of GRB 230307A: Signature of an Evolving Radiative Efficiency in Magnetar Wind Dissipation?}
\author[0000-0002-1766-6947]{Shu-Qing Zhong}
\affil{School of Astronomy and Space Science, University of Science and Technology of China, Hefei 230026, China; daizg@ustc.edu.cn}
\author[0000-0002-8391-5980]{Long Li}
\affil{School of Astronomy and Space Science, University of Science and Technology of China, Hefei 230026, China; daizg@ustc.edu.cn}
\author[0000-0002-4304-2759]{Di Xiao}
\affil{Purple Mountain Observatory, Chinese Academy of Sciences, Nanjing 210023, China}
\author[0000-0002-9615-1481]{Hui Sun}
\affil{Key Laboratory of Space Astronomy and Technology, National Astronomical Observatories, Chinese Academy of Sciences, Beijing 100012, China}
\author[0000-0003-4111-5958]{Bin-Bin Zhang}
\affil{School of Astronomy and Space Science, Nanjing University, Nanjing 210023, China}
\author[0000-0002-7835-8585]{Zi-Gao Dai}
\affil{School of Astronomy and Space Science, University of Science and Technology of China, Hefei 230026, China; daizg@ustc.edu.cn}

\begin{abstract}
Very recently, a particularly long gamma-ray burst (GRB) 230307A was reported and
proposed to originate from a compact binary merger based on its host galaxy property, kilonova, and heavy elements.
More intriguingly, 
a very early plateau followed by a rapid decline in soft X-ray band was detected in its light curve by the Lobster Eye Imager for Astronomy, 
indicating strong evidence of the existence of a magnetar as the merger product.
This work explores that the Magnetar Wind Internal Gradual MAgnetic Dissipation (MIGMAD) model, in which the radiative efficiency evolves over time, successfully fits it to the observed data. Our results reinforce the notion that the X-ray plateau serves as a powerful indicator of a magnetar and imply that an evolving efficiency is likely to be a common feature in X-ray plateaus of GRB afterglows.
In addition, we also discuss the explanations for the prompt emission, GRB afterglows, as well as kilonova, and predict possible kilonova afterglows in a magnetar central engine.

\end{abstract}

\keywords{Compact binary stars (283); Gamma-ray bursts (629); Magnetars (992)}

\section{Introduction}
\label{sec:introduction}
According to the observed bimodal distribution of duration \citep{kou93}, it is widely accepted that gamma-ray bursts (GRBs) are generally classified as short GRBs (SGRBs) originating from the merger of a binary compact object, and long GRBs (LGRBs) invoking the core collapse of a massive star. However, this classification is unreliable for a few apparently SGRBs with a collapsar origin, e.g., GRBs 090426 \citep{leve10} and 200826A \citep{ahu21,zhang21}, and some apparently LGRBs with a merger origin, e.g., GRBs 060614 \citep{gal06,della06,fyn06,geh06}, 211227A \citep{lv22}, 211211A\footnote{Recent simulations \citep{sie19} and case study \citep{lil23} suggested that a ``collapsar'' supernova associated with a LGRB may also produce r-process elements which likely give rise to a kilonova-like component, but \cite{bar23} showed that a collapsar origin for GRB 211211A is just barely possible because of the requirements of high kinetic energy and an unexpected pattern of $^{56}$Ni enrichment.} \citep{ras22,yang22,tro22,zhang22,mei22}, and the most recent case GRB 230307A \citep{levan23,sun23,gill23,yang23}.   

GRB 230307A is most likely hosted by a bright spiral galaxy at $z\sim0.065$, contains a kilonova similar to AT2017gfo, and has a spectroscopic evidence of tellurium emission line \citep{levan23,gill23,yang23}.
Its own prompt emission contains a long-duration main burst with two breaks in decay phase \citep{sun23} and a possible precursor \citep{dich23}, 
followed by optical and radio afterglows \citep{levan23} as well as a soft X-ray plateau in afterglow light curves \citep{sun23}.

X-ray plateaus in GRB afterglows are usually classified as ``external plateaus'' and ``internal plateaus'' \citep[e.g.,][]{tro07,lv14}. External plateaus are interpreted as the decelerating external forward shock by introducing: continuous energy injection from the central engine \citep{dai98a,dai98b,zhang01,rees98,sari00,zhang06}, a two-component jet \citep{racu08}, 
a reverse shock \citep{uhm07,gen07}, evolution of microphysical parameters \citep{ioka06}, or slightly misaligned viewing angle to the GRB jet \citep{eich06,ben20}.
Internal plateaus result from internal energy dissipation in the wind of a central magnetar \citep{cor90,usov94} 
or in the fallback accretion of the stellar envelope into a black hole \citep[BH;][]{kumar08a,kumar08b}. 
The very early soft X-ray plateau followed by a decline phase with a slope of $-2$ in GRB 230307A afterglows, detected by the Lobster Eye Imager for Astronomy \citep[LEIA;][]{zhangc22,ling23}, shows a chromatic behavior with the optical afterglows from \cite{levan23}.
Moreover, its spectrum exhibits less significant evolution within the first 100 s \citep{sun23}. 
The former feature rules out the external forward shock by introducing energy injection, evolution of microphysical parameters, or slightly misaligned viewing angle to the GRB jet. 
While the latter feature usually disfavors the external forward shock by introducing a two-component jet or a reverse shock. 
As a result, these features suggest that the soft X-ray plateau is most likely to belong to an internal plateau: 
the said plateau dominated by internal dissipation emission but with a slope of $<3$ in decline phase \citep[see Conclusions and Discussion in][]{lv14}.
Internal X-ray plateaus in SGRB afterglows are typically attributed to internal dissipation of the magnetar wind rather than the fallback accretion into a BH \footnote{The fallback accretion of the stellar envelope into a BH can interpret internal X-ray plateaus in LGRBs \citep{kumar08a,kumar08b}, but it usually cannot explain internal X-ray plateaus in SGRBs since the required fallback mass is too large for SGRBs with a merger origin.}.

\begin{figure*}
	\centering
	\includegraphics[width=0.48\textwidth]{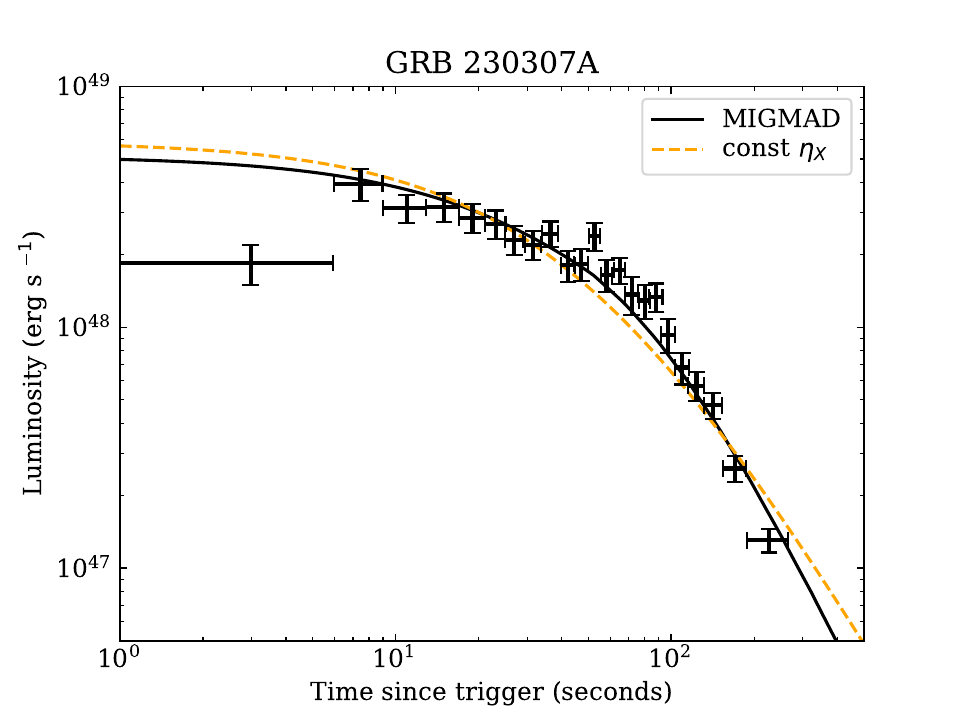}
	\includegraphics[width=0.48\textwidth]{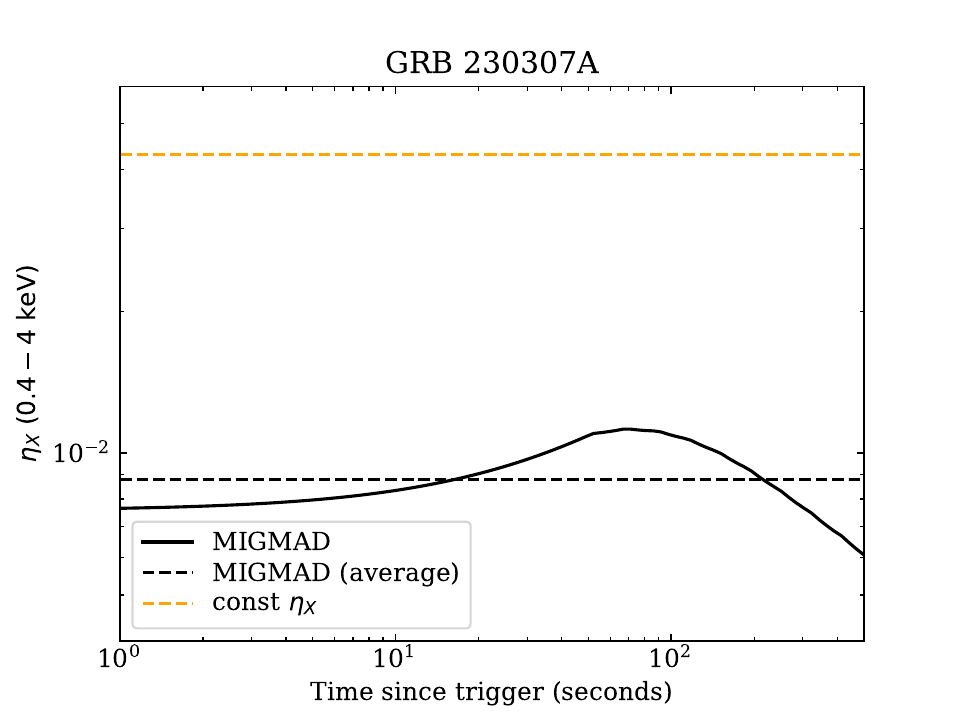}
	\caption{Left panel: The best-fit results with an evolving radiative efficiency via MIGMAD model (black solid line) and with a constant efficiency (orange dashed line). Right panel: The corresponding evolving and constant efficiencies. The time-averaged value $<\log_{10}(\eta_X)>=-2.06$ of the evolving efficiency is plotted with a black dashed line.}
	\label{fig:GRB}
\end{figure*}

The X-ray plateau in GRB 230307A afterglows, which has a similar shape with the X-ray transient CDF-S XT2 \citep{xue19}, therefore is naturally ascribed to internal dissipation of the magnetar wind.
While the numerical modeling with a constant radiative efficiency during the wind dissipation successfully reproduces the X-ray data, it is not good enough especially for the decay phase, as shown by Figure 4 in \cite{sun23}.
This may suggest an evolving efficiency in the wind dissipation, which is exactly the focus of the paper.

\begin{deluxetable}{ccccccccc}
	\label{tab:paras}
	\tabletypesize{\scriptsize}
	\tablecaption{The Best-fit Results for the X-ray Plateaus of GRB 230307A}
	\tablehead{
		\colhead{Parameters} &
		\multicolumn{4}{c}{Values}
	}
	\startdata
	\object{}  &  \multicolumn{2}{c}{MIGMAD} &  \multicolumn{2}{c}{Const $\eta_X$}  \\
	\object{}  &  Priors & Results & Priors & Results \\
	\hline
	\object{$\log_{10}(\eta_X)$}  &  &   Evolving$^a$ & [$-6.0$, $0.0$] & $-1.61^{+0.41}_{-0.42}$  \\
	\object{$\log_{10}(B_{\rm p}/{\rm G})$} & [12.0, 17.0] &  $16.04^{+0.05}_{-0.04}$ & [12.0, 17.0]  & $16.23^{+0.23}_{-0.21}$  \\
	\object{$\log_{10}(P_0/{\rm s})$} & [$-3.1$, $-1.0$] &  $-3.03^{+0.05}_{-0.04}$ &  [$-3.1$, $-1.0$] & $-2.81^{+0.22}_{-0.20}$  \\
	\object{$\log_{10}(\lambda/\epsilon)$} &  [7.0, 10.0] &  $7.69^{+0.23}_{-0.18}$ & &   \\
	\object{$\epsilon_e$}  & [0.001, 0.5] & $0.41^{+0.05}_{-0.07}$ &  &  \\
	\object{$\log_{10}(\Gamma_{\infty})$}  & [1.0, 4.0] & $3.25^{+0.04}_{-0.05}$ & &   \\
	\object{$p$}   & [1.0, 4.0] &  $2.74^{+0.15}_{-0.16}$ &    \\
	\object{$\chi^2$/dof}  &  & 33/16=2.1  & & 73/19=3.8  \\
	\object{ BIC value$^b$}  &  & 52  & & 82   \\
	\enddata
	\tablenotetext{a}{ The efficiency in the MIGMAD model is not a free parameter, which relies on other free parameters as well as evolves with time and has a time-averaged value $<\log_{10}(\eta_X)>=-2.06$ from the MCMC results.}
	\tablenotetext{b}{ The Bayesian information criterion (BIC) value is calculated by ${\rm BIC}=\ln(n)k-2\ln(L)$ in which $n$ is the number of data point, $k$ is the number of model parameters, and $L$ is the likelihood function.}	
\end{deluxetable}

\begin{figure}
	\centering
	\includegraphics[width=0.5\textwidth]{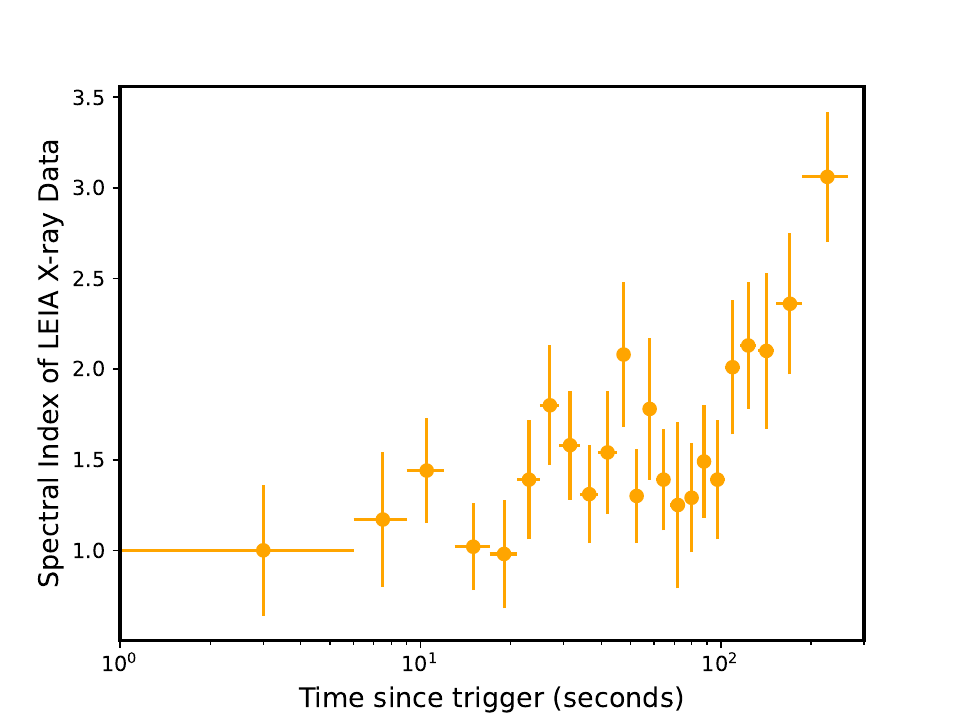}
	\caption{The spectral evolution of LEIA X-ray data plotted from the Extended Data Table 1 of \cite{sun23}.}
	\label{fig:spec_index}
\end{figure}

\section{Magnetar Wind Dissipation with an Evolving Radiative Efficiency}
\label{sec:wind}
In the modeling result for the X-ray plateau by the Markov chain Monte Carlo (MCMC) algorithm in \cite{sun23}, the gravitational wave (GW) torque to the magnetar spin-down is found to be negligible, so the spin-down is dominated by the magnetic dipole torque.   
In this case, the magnetar rotational energy loss rate can be written as \citep[e.g.,][]{sha83}
\begin{equation}
	\dot{E}=I \Omega \dot{\Omega}=-\frac{B_{\mathrm{p}}^2 R^6 \Omega^4}{6 c^3},
	\label{eq:E_dot}
\end{equation}
where $B_{\rm p}$, $\Omega=2\pi/P$, and $R$ are the surface dipole magnetic field strength, spin angular velocity, and radius of the magnetar, $P$ is its spin period, $c$ is the speed of light, $I=\frac{2}{5}MR^2$ is its moment of inertia in which $M$ is the mass, throughout the paper $R=10^6$ cm and $M=2.0M_{\odot}$ are adopted.
One can solve above equation and obtain the form of $\Omega(t)$.
Further the observed X-ray plateau luminosity can be obtained from the magnetic dipole luminosity after introducing a radiative efficiency $\eta_X$
\begin{equation}
	L_{\mathrm{X}}(t)=\eta_X L_{\mathrm{sd}}=\eta_X \frac{B_{\mathrm{p}}^2 R^6 \Omega^4(t)}{6 c^3}.
\end{equation}

\begin{figure*}
	\centering
	\includegraphics[width=1\textwidth]{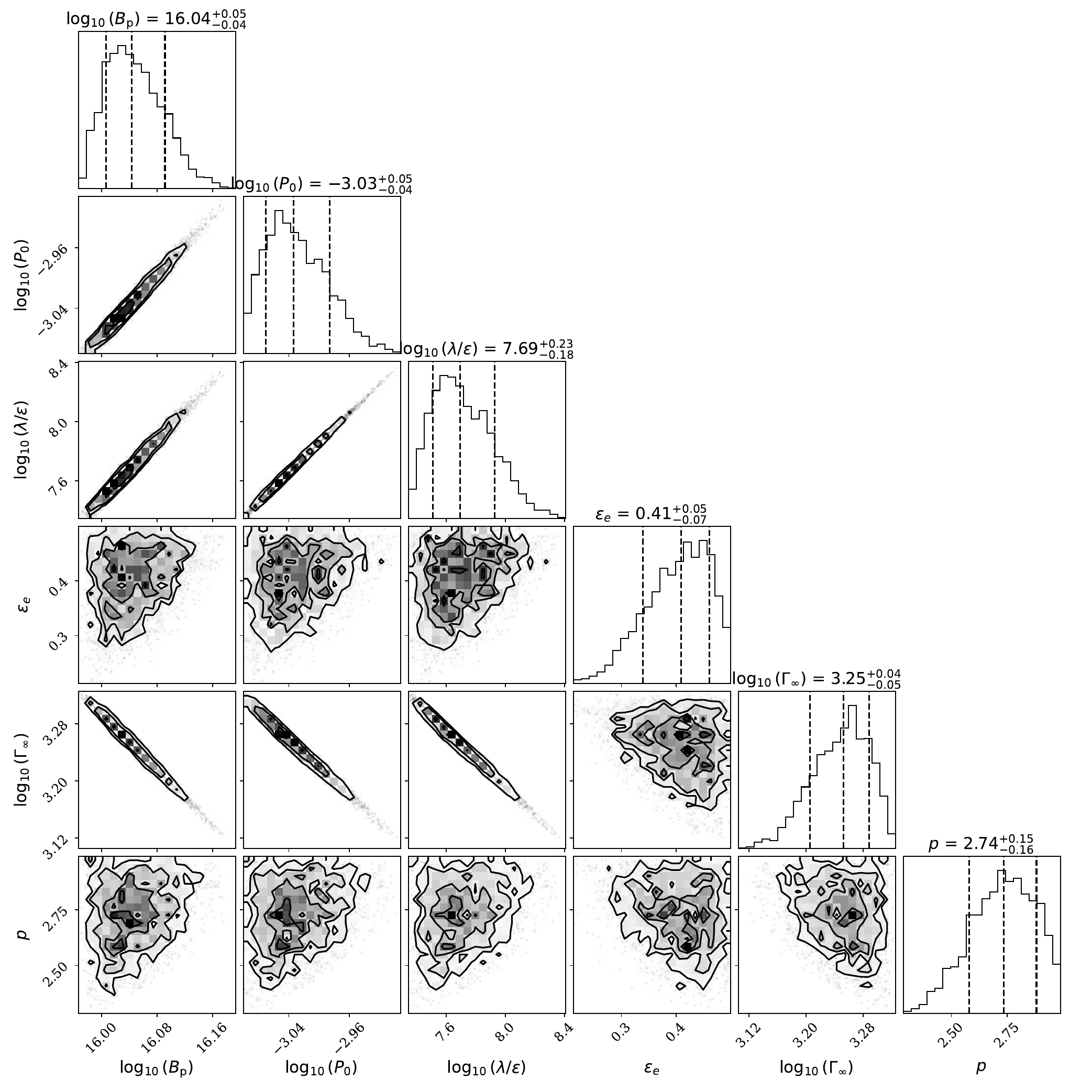}
	\caption{The corner plot for the X-ray plateau of GRB 230307A.}
	\label{fig:corner_GRB}
\end{figure*}

Considering the radiative process from the magnetar spin-down to the observed X-ray radiation via the magnetar wind internal gradual magnetic dissipation (MIGMAD) model \citep{ben17,xiao17,xiao18}, as done in \cite{xiao19a} for X-ray plateaus in a few GRB afterglows and in \cite{xiao19b} for CDF-S XT2, 
the efficiency strongly depends on the injected luminosity and the wind's saturation Lorentz factor. 
Therefore, the radiative efficiency should evolve with time rather than being a constant.
The basic picture of the MIGMAD model is illuminated here.
A newborn millisecond magnetar serves as the central engine and powers an initial Poynting-flux dominated outflow with a ``striped wind'' magnetic field configuration. Then the magnetic energy in the outflow is dissipated via internal gradual reconnection process and partially converted to high-energy emission that consists of a thermalized component and a non-thermal component from synchrotron radiation of accelerated electrons. There are several free parameters in the model: the surface dipole magnetic field strength $B_{\rm p}$, the initial spin period $P_0$ of the magnetar,  the wavelength of the
magnetic field in the striped wind configuration $\lambda$ \citep{cor90,spr01,dren02a,dren02b}, the ratio of reconnection velocity to the
speed of light $\epsilon$ \citep{guo15,liu15}, the fraction of dissipated energy into electrons $\epsilon_e$, 
the bulk Lorentz factor of the wind at the saturation radius $\Gamma_{\infty}$, 
and the electron distribution with an index $p$ accelerated by reconnection \citep{sir14,guo15,kag15,wer16}. For further details one can refer to \cite{xiao18}.
Through the MCMC algorithm, we have a best reproduction for the observed LEIA X-ray data with parameter values $\log_{10}(B_{\rm p}/{\rm G})=16.04^{+0.05}_{-0.04}$, $\log_{10}(P_0/{\rm s})=-3.03^{+0.05}_{-0.04}$, $\log_{10}(\lambda/\epsilon)=7.69^{+0.23}_{-0.18}$, $\epsilon_e=0.41^{+0.05}_{-0.07}$, $\log_{10}(\Gamma_{\infty})=3.25^{+0.04}_{-0.05}$, 
and $p=2.74^{+0.15}_{-0.16}$, which are listed in Table \ref{tab:paras} and plotted in Figure \ref{fig:corner_GRB} with a corner.
The corner plot shows some degenerate parameters. For example, $B_{\rm p}$ and $P_0$ is degenerate because of the total luminosity of the wind $L_{\rm sd}$ (Equation (\ref{eq:E_dot})), $P_0$ and $\lambda/\epsilon$ because of the typical wavelength of the field $\lambda=cP$ in the striped wind configuration \citep{cor90,spr01,dren02a,dren02b}, as well as $\lambda/\epsilon$ and $\Gamma_{\infty}$ because of the bulk Lorentz factor of the wind at the saturation radius 
given by $r_{\infty}=\lambda \Gamma_{\infty}^2/(6\epsilon)$ \citep{ben17}. 
From the MCMC results, the model prediction of spectral evolution from $F_{\nu}\propto \nu^{-0.5}$ to $F_{\nu}\propto \nu^{-(p-1)/2}$ \citep{xiao19a} obtained by the electron distribution index $p\sim2.74$ generally matches the transition of the observed X-ray photon spectral index from an average value of $\sim1.4$ before $\sim100$ s to $\sim2.1$ after $\sim100$ s, see Figure \ref{fig:spec_index} plotted from the Extended Data Table 1 of \cite{sun23}.
Note that the efficiency in the MIGMAD model is not a free parameter, which relies on other free parameters and evolves with time. 
It has a time-averaged value $<\log_{10}(\eta_X)>=-2.06$, smaller than the value $\log_{10}(\eta_X)=-1.61$ in the constant efficiency scenario.

The fitting results are exhibited in Figure \ref{fig:GRB}, in which the best fit with a constant efficiency is also displayed as a comparison.
As one can see, the best fit with an evolving efficiency very well reproduces the observed data. 
It is much better than the fit with a constant efficiency, especially for the decay phase of the X-ray data. 
This also can be directly seen due to the goodness comparison between these two fits with an evolving efficiency and with a constant efficiency, from both the $\chi^2$/dof difference (2.1 vs. 3.8) and the BIC difference ($\Delta {\rm BIC}=-30$), see Table \ref{tab:paras}.
The result comparison with and without an evolving efficiency suggests that a specific wind dissipation process with evolving efficiency 
should present in the conversion from the magnetar spin-down power to the observed X-ray radiation. 
The requirement for such an evolving efficiency is likely to be prevalent in X-ray plateaus of GRB afterglows. 

\begin{figure*}
	\centering
	\includegraphics[width=1.0\textwidth]{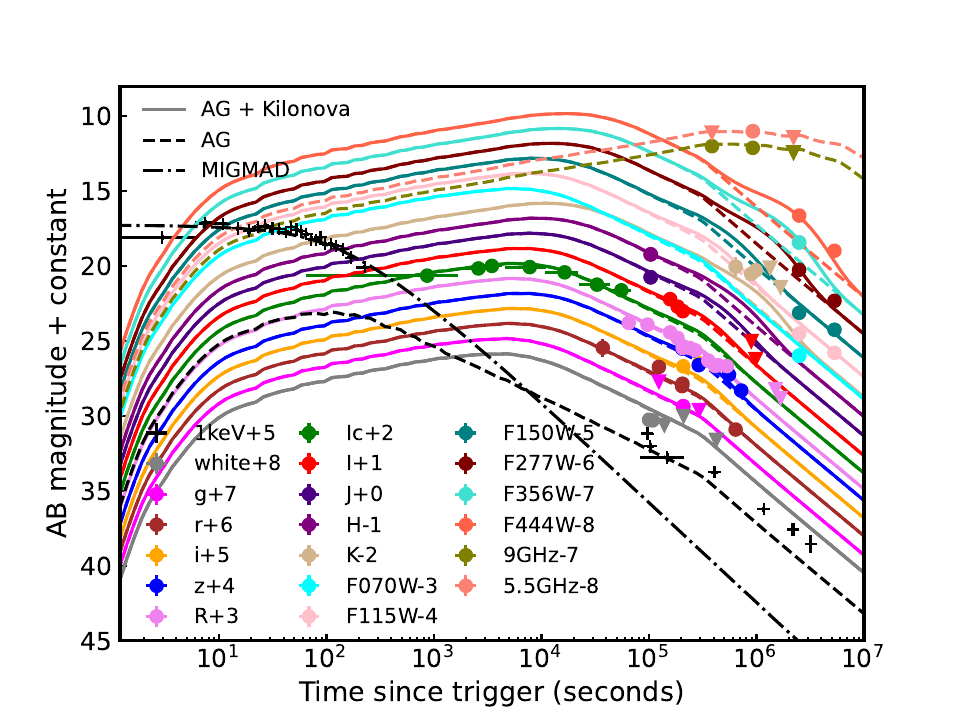}
	\caption{The best joint fitting for the multi-wavelength afterglow and kilonova light curves of GRB 230307A by combining the standard external forward shock model and the kilonova model. 
	The early X-ray (1keV) data before 300 s is scaled from the LEIA data and the late is scaled from the Swift XRT, XMM-Newton, and Chandra data. 
	The early X-ray data as a plateau is fitted by the MIGMAD model (black dashed-dotted line), same as the left panel of Figure \ref{fig:GRB}.
	The data points denoted by an inverted triangle are upper limits. The GRB afterglow (AG) components are denoted by dashed lines.}
	\label{fig:afterglows}
\end{figure*}

\section{Prompt Emission, GRB Afterglows, Kilonova, and Kilonova Afterglows in a Magnetar Central Engine}
\label{sec:prompt}
{\em Prompt Emission.} In general, there are two types of models to generate a prompt emission if the central engine is a magnetar. 
One extract uniform rotation of magnetar and invoke an internal dissipation of magnetar wind \citep{usov92,met11,ben17}, 
the other extract its differential rotation and invoke magnetic bubbles \citep{klu98,dai98b,spr99,rud00,dai06,zhong23}.
In this work, we hold the opinion that the prompt emission of GRB 230307A is likely produced by the differential-rotation-induced magnetic bubbles.
This is because: (1) we have considered that the X-ray plateau of GRB 230307A is relevant to an internal dissipation of magnetar wind; 
(2) the process to the prompt emission should be different from that to the X-ray plateau, 
which is supported by the joint spectral fittings and the light curve behaviors of GECAM and LEIA data, see \cite{sun23}. 
In this case, the information of the prompt emission from the magnetar differential rotation such as its energy and duration is not required to compare with the energy reservoir and duration of the magnetar uniform rotation but its differential rotation. As known by the observed properties of the prompt emission \citep[see Table 1 in][]{sun23} and the fitting for the GRB multi-wavelength afterglows below (Table \ref{tab:afterglows}), the total isotropic energy of the jet is $E_{\rm jet, iso}=E_{\gamma, \rm iso}+E_{\rm K, \rm iso}=3.08\times10^{52}+5.62\times10^{50}\approx 3.14\times10^{52}$ erg\footnote{The extremely high radiative efficiency $\eta_{\gamma}\approx98\%$ of the jet is consistent with the view that the jet is Poynting-flux dominated, suggested by \cite{yi23} and \cite{du24}. This efficiency and its inferred Poynting-flux dominated jet are naturally consistent with the jet origin from differential-rotation-induced magnetic bubbles suggested in our work.}. The maximum kinetic energy of a differentially rotating newly-born magnetar up to $3\times10^{53}$ erg \citep{cook94} is certainly adequate for the jet energetics. The uniform rotation energy of the magnetar $E_{\Omega}=\frac{1}{2} I \left(\frac{2\pi}{P_0}\right)^2 \approx 2\times10^{52}$ erg inferred from Table \ref{tab:paras} is also certainly adequate for the X-ray plateau energetics \citep[$2.47\times10^{50}$ erg in Table 1 of][]{sun23}, with a time-averaged radiative efficiency $<\log_{10}(\eta_X)>=-2.06$. This small efficiency implies that the lost of the uniform roration energy in the X-ray plateau is nearly negligible. As will be shown in the following, the rest of the uniform rotation energy should be poured into the kilonova ejecta and ultimately the circumburst medium (CSM), and make its radio afterglow much brighter.

{\em GRB Afterglows and Kilonova.} In a magnetar central engine, the transition from internal to external dissipation in the X-ray afterglow should occur post the LEIA data but prior to the Swift XRT data, as shown by the best joint fitting for the multi-wavelength GRB afterglow and kilonova light curves in Figure \ref{fig:afterglows}. Within which the early X-ray (1keV) data before 300 s are scaled from the LEIA \citep{sun23} while the late are collected from the Swift XRT, XMM-Newton, and Chandra \citep{yang23}, the optical and radio data are collected from \cite{levan23}. The best joint fitting is obtained by the MCMC technique, within the framework of combining the standard external forward shock model \citep{sari98,huang99} and the kilonova model used in \cite{xiao17b}. Its corresponding parameter results are listed in Table \ref{tab:afterglows}. Within which there are seven free parameters for the standard external forward shock model: the jet half-opening angle $\theta_{\rm j}$, the isotropic kinetic energy $E_{\rm K,iso}$,
the initial Lorentz factor $\Gamma_0$,
the ratio of shock energy to the magnetic field $\epsilon_B$, the ratio of shock energy to the electron $\epsilon_{e,\rm f}$, the interstellar medium (ISM) density $n_{\rm ISM}$,
and the power-law index of the electron distribution $p_{\rm f}$, 
while four free parameters for the kilonova model: the velocity $v_{\rm ej}$, the polar opening angle $\theta_{\rm ej}$, the mass $M_{\rm ej}$, and the opacity $\kappa$ of the ejecta.

{\em Kilonova Afterglows.} Due to the significantly small uniform rotation energy lost in the X-ray plateau, the energy poured into the kilonova ejecta can be the total uniform rotation energy $E_{\Omega}$. According to the estimate of \cite{met14}, the kilonova afterglow in radio band $\nu=1.4$ GHz should have a peak flux density given by
\begin{equation}
	\begin{aligned}
	F_{\rm peak} \approx & 3~{\rm mJy}\left(\frac{E_{\Omega}}{10^{52} \rm erg}\right)\left(\frac{n_{\rm ISM}} {{\rm cm}^{-3}}\right)^{0.83}\left(\frac{\epsilon_{B,\rm ej}}{0.1}\right)^{0.83} \\
	&\times\left(\frac{\epsilon_{e, \rm ej}}{0.1}\right)^{1.3} \beta_0^{2.3} d_{28}^{-2}
	\label{eq:Flux}
\end{aligned}
\end{equation}
at the deceleration time 
\begin{equation}
	t_{\rm dec} \approx 1.2~{\rm yr}\left(\frac{E_{\Omega}}{10^{52} {\rm erg}}\right)^{1 / 3}\left(\frac{n_{\rm ISM}}{{\rm cm}^{-3}}\right)^{-1 / 3} \beta_0^{-5 / 3},
	\label{eq:t_dec}
\end{equation}
where $\epsilon_{B,\rm ej}$ and $\epsilon_{e, \rm ej}$ are the ratios of shock energy to the magnetic field and to the electron in the ejecta-related shock, $d=10^{28} d_{28}$ cm is the luminosity distance, and $\beta_0\sim 1.0\left(\frac{E_{\Omega}}{10^{52} \rm erg}\right)^{1 / 2}\left(\frac{M_{\rm ej}}{10^{-2} M_{\odot}}\right)^{-1 / 2}$ is the accelerated velocity of the matter in which the magnetar wind couples its energy to the ejecta.
Since $E_{\Omega}$, $n_{\rm ISM}$, and $M_{\rm ej}$ have been obtained (see Table \ref{tab:afterglows}), if given $\epsilon_{B,\rm ej}=0.1$ and $\epsilon_{e, \rm ej}=0.1$, one would have $F_{\rm peak}\approx0.66$ mJy at $t_{\rm dec}\approx15$ yr. This predicted late-time radio emission may be used as a critical way to test whether the central engine of GRB 230307A is a millisecond magnetar.

\begin{deluxetable}{ccccc}
	\label{tab:afterglows}
	\tablecaption{The Best Parameter Distributions for Jointly Modeling the Multi-wavelength Afterglows and Kilonova of GRB 230307A by Combining the Standard External Forward Shock Model and the Kilonova Model}
	\tablehead{
		\colhead{Parameters} &
		\colhead{Priors} &
		\colhead{Values}
	}
	\startdata
	\object{Standard Afterglow Model} & &     \\
	\hline
	\object{$\theta_{\rm j}$ (rad)} & [0.001, 0.5] &  $0.08^{+0.02}_{-0.01}$    \\
	\object{$\log_{10}(E_{\rm K,iso}/{\rm erg})$} & [48.0, 55.0] &  $50.76^{+0.26}_{-0.28}$     \\
	\object{$\log_{10}(\Gamma_0)$} & [2.0, 4.0] &  $2.77^{+0.28}_{-0.27}$     \\
	\object{$\log_{10}(n_{\rm ISM}/{\rm cm}^{-3}$)}  & [$-5.0$, $-2.0$] &  $-4.43^{+0.42}_{-0.39}$     \\
	\object{$\log_{10}(\epsilon_B)$}  & [$-7.0$, $-0.5$] &    $-0.87^{+0.25}_{-0.54}$   \\
	\object{$\log_{10}(\epsilon_{e,\rm f})$}   & [$-6.0$, $-0.5$]  &    $-0.59^{+0.06}_{-0.07}$   \\
	\object{$p_{\rm f}$}   & [2.0, 3.0] &    $2.83^{+0.05}_{-0.03}$   \\
	\hline
	\object{Kilonova Model} & &     \\
	\object{$\log_{10}(v_{\rm ej}/c)$}   & [$-2.0$, $-0.3$]  &    $-1.27^{+0.01}_{-0.01}$   \\
	\object{$\theta_{\rm ej}$ (rad)}   & [$\pi/8$, $3\pi/8$]  &    $0.75^{+0.27}_{-0.25}$   \\
	\object{$\log_{10}(M_{\rm ej}/M_{\odot})$}   & [$-3.0$, $-1.3$]  &    $-2.27^{+0.03}_{-0.04}$   \\
	\object{$\log_{10}(\kappa/{\rm cm^2~g^{-1}})$}   & [$-1.0$, $2.0$]  &    $1.31^{+0.14}_{-0.16}$   \\
	\enddata
\end{deluxetable}

\section{Summary and Discussion}
\label{sec:summary}
We have presented numerical fittings for the X-ray plateau of GRB 230307A with and without MIGMAD model. 
The comparative fitting results imply a signature of an evolving efficiency in the magnetar wind dissipation from the magnetar spin-down power to the observed X-ray plateau.  
Moreover, such an evolving efficiency in magnetar wind dissipation is likely to be ubiquitous in those X-ray plateaus of afterglows for either SGRBs or LGRBs have been fitted by many authors using an empirical smoothly broken power-law function \citep[e.g.,][]{row13,lv14,lv15,tang19,zou19,zou21}. 
We have also discussed the explanations for the prompt emission, GRB afterglows, as well as kilonova, and also predicted possible kilonova afterglows in a magnetar central engine.

In this work, one caveat related to the X-ray data should be kept in mind, that is, 
the first data point is lower than the second one.
This can be explained by, during the initial ten seconds after the birth of proto-magnetar, the magnetar wind that is baryon-loaded due to the strong neutrino heating and thus has a small $\Gamma_{\infty}$, as already noticed in \cite{met11}.
This therefore leads to an inefficient dissipation of the magnetar wind and a very low value of  $\eta_X$ via the MIGMAD model.

Another caveat is the possibility that magnetar wind powers internal X-ray plateaus in non-collapsar GRBs is challenged by the distributions of energies and durations of internal plateaus \citep[see][]{ben21}, the ought-to-be brighter kilonova boosted by magnetar wind \citep[mergernova;][]{yu13,gao13}, 
as well as by the possible absent of forthcoming late-time radio emission of SGRBs due to the interaction between the merger ejecta and CSM \citep{met14,fong16,hor16,ricci21}. 
The second inconsistency between the mergernova expectation and the not-too-bright kilonova of GRB 230307A \citep{levan23,yang23} could be mediated by a quite low heating efficiency \citep{ai22}. But a more recent work by \cite{wang24} discussed a counter-argument to this low heating efficiency suggestion. While the last may be mediated by a much lower circumburst median density \citep{met14}, a relatively lower $\epsilon_{B,\rm ej}$ \citep{liu20}, or an incomplete sweeping on the interstellar medium of kilonova ejecta \citep{lis23}, though we have predicted a bright late-time radio emission peaking around 0.66 mJy at $\sim 15$ yr in Section \ref{sec:prompt}.

While most of current observations seem to favor a compact binary merger for the progenitor of GRB 230307A, it is still on debate that what the compact binary is.
If the final object is a magnetar as required in our work for GRB 230307A, similar to GRB 211211A, 
the progenitor system could only be neutron star$-$neutron star (NS$-$NS) or neutron star$-$white dwarf binaries\citep[NS$-$WD;][]{yang22,zhong23}.
For an NS$-$NS merger, however, how to explain the particularly long duration of GRB 230307A may be a challenge.
While for an NS$-$WD merger, the production of a kilonova is an open question. 

\cite{yang22} suggested that the NS might merge into the center of the WD and induce the collapse of the WD, 
then kilonova could be produced by the ejecta consisting of a neutron-poor component and a neutron-rich component. 
The former component is from the materials of the WD neutronized and mixed with the materials of the pre-merger NS during the WD collapsing,  while the latter is from the disk wind after a proto-magnetar forms. 
This idea is novel and its key assumption is that the WD is not totally disrupted by the NS. 
However, this assumption is contrast to the conventional view that the WD is completely tidally disrupted by the NS
\citep{met12,fer13,mar16,bob17,bob22,fer19,zen19,zen20,kal23,mor24}.
Moreover, whether the pre-merger NS can induce the collapse of the WD is unknown. 

In the model of the tidally-disrupted NS$-$WD merger of \cite{zhong23} in which the WD is completely tidally disrupted by the NS, 
the kilonova-like emission of GRB 211211A might actually be a fast-evolving transient like SN 2005ek \citep{dro13} or SN 2018kzr \citep{mc19,gil20} 
powered by $^{56}$Ni radioactive decay 
adding an energy injection from central engine. 
While for GRB 230307A, r-process heavy elements are found in the kilonova-like emission. 
In a tidally-disrupted NS$-$WD merger, these r-process elements could be produced by the neutron-rich material away from the newly formed postmerger magnetar. For which the neutron-rich material is accelerated by a magnetic jet. This picture should be similar to that summarized in \cite{sie22} for magnetorotational or ``jet-driven'' core-collapse supernovae. That is, in a competition between neutrino irradiation and acceleration by the magnetic jet, a fraction of the ejecta may evade the strong protonizing neutrino irradiation from the postmerger hot magnetar to remain relatively neutron-rich. Production of r-process elements beyond the second peak requires a magnetar-strength magnetic field and a millisecond rotation period to trigger a prompt jet-like ejection of neutron-rich matter \citep{win12,nis17,hal18,mos18,rei21}. 
These requirements seem to be met in the tidally-disrupted NS$-$WD merger model of \cite{zhong23}. 
The possibility that the tidally-disrupted NS$-$WD merger model accounts for all the observations of GRB 230307A, particularly the r-process heavy elements, will be explored elsewhere.

\acknowledgments
We are very grateful to thank the referee for the careful and thoughtful suggestions that have helped us improve this manuscript substantially.
We thank Ruo-Yu Liu for helpful discussions.
This work is supported by the National SKA Program of China (grant No. 2020SKA0120300) 
and National Natural Science Foundation of China (grant No. 12393812).  
S.Q.Z. acknowledges support from the National Natural Science Foundation
of China (grant No. 12247144) and China Postdoctoral Science Foundation (grant Nos. 2021TQ0325 and 2022M723060).
L.L. is supported by the National Natural Science Foundation of China (grant No. 12303050), China Postdoctoral Science Foundation (grant No. 2023M743397), 
and Fundamental Research Funds for the Central Universities. 
D.X. is supported by the National Natural Science Foundation of China (grant No. 12373052).
H.S. is supported by the National Natural Science Foundation of China (grant No. 12103065).
BBZ acknowledges the support by the National Key Research and Development Programs of China (2022YFF0711404, 2022SKA0130102), the National SKA Program of China (2022SKA0130100), the National Natural Science Foundation of China (Grant Nos. 11833003, U2038105, 12121003), the science research grants from the China Manned Space Project with NO.CMS-CSST-2021-B11, the Fundamental Research Funds for the Central Universities, and the Program for Innovative Talents and Entrepreneur in Jiangsu.



\end{document}